\documentclass[12pt]{article}
\usepackage{graphicx}
\title{Propagation, breathing and transition of matter-wave packet trains}
\author{Wenhua Hai$^{*a,b}$, \  Chaohong Lee$^{b}$, \ Guishu Chong$^a$
\\ $^a$Department of Physics, Hunan Normal University, Changsha
410081, China \\  $^b$Laboratory of Magnetic Resonance and Atomic
and Molecular Physics, \\ Wuhan Institute of Physics and
Mathematics, Chinese Academy of Sciences,
\\ Wuhan 430071, China}
\date{}
\begin{document}
\maketitle
\begin{abstract}
We find a set of new exact solutions of a quantum harmonic
oscillator, which describes some wave-packet trains with average
energy being proportional to both the quantum level and classical
energy of the oscillator. Center of the wave-packet trains may
oscillate like a classical harmonic oscillator of frequency
$\omega$. Width and highness of the trains may change
simultaneously with frequency $2 \omega $ as an array of
breathers. Under some perturbations the wave-packet trains could
transit between the states of different quantum numbers. We
demonstrate analytically and numerically that the wave-packet
trains can be strictly fitted to the matter-wave soliton trains
observed by Strecher et al. and reported in Nature 417, 150(2002).
When the wave-packets breathe with greater amplitudes, they show
periodic collapse and revival of the matter-wave.

\end{abstract}
\footnotetext{$^*$ Email address: adcve@public.cs.hn.cn}
\newpage

As an elementary equation of quantum mechanics the Schr$\ddot
o$dinger equation is a linearly partial differential one of the
second order with variable coefficients \cite{Landau}-\cite{Zeng}.
We well known that this equation cannot be exactly solved yet for
most physically interesting systems, except a few systems with
separation of variables such as hydrogen atom, harmonic oscillator
and rigid rotator \cite{Steeb}, \cite{Hai1}. The quantum states
described by its solutions with inseparable space-time variables
are very important but difficult to find. The coherent state of a
harmonic oscillator is a nice example of such states
\cite{Schrodinger}-\cite{Howard}. To seek new inseparable exact
solutions of a Schr$\ddot o$dinger equation and to physically
realize them are our main motivations in this paper.

The preparation and measurement of quantum states are very hard
even impossible for some microscopic systems \cite{Royer}-
\cite{Kurtsiefer}. Compared to this the physical realization and
detection of the macroscopic and mesoscopic quantum states may be
easier sometimes \cite{Monroe}, \cite{Brune}. The Bose-Einstein
condensate (BEC) just supplies such a macroscopic quantum system
\cite{Anderson}-\cite{Andrews}, which can be use to test the
Schr$\ddot o$dinger quantum mechanics. For example, a harmonically
confined BEC can be identified as a perturbed quantum harmonic
oscillator, when the interatomic interaction is weak enough
\cite{Dalfovo}, \cite{Leggett}. The weak atom-atom interactions
were due to the small atomic samples \cite{Wright}, \cite{Wright2}
and short $s$-wave scattering length $|a|$ \cite{Strecher},
\cite{Khaykovich} that can be controlled by the Feshbach resonance
\cite{Tiesinga}, \cite{Inouye}.

A quite interesting phenomenon was experimentally observed that
the weakly interacting BEC appears solitonlike behavior
\cite{Strecher}, \cite{Khaykovich}. Although this was
approximately explained by using nonlinear interaction
\cite{Khawaja}, \cite{Denschlag}, Strecher and coworkers said that
"non-interacting solitons $\cdots$ would be expected to pass
through one another" in their experiment. The solitonlike behavior
is also explored for ideal BEC gas with $a=0$ in Khaykovich's
experiment. Another important fact is the finding of collapse and
revival of the macroscopic matter-wave packets \cite{Greiner},
\cite{Sackett}, which can occur for very weak interaction with
atomic samples composed of a few thousand particles \cite{Wright}.
In this paper we shall report a set of new exact solutions of a
quantum harmonic oscillator and the corresponding macroscopic
quantum level. By using them we demonstrate that, analytically and
numerically, the wave-packet trains governed by these exact
solutions can be strictly fitted to the matter-wave soliton trains
found by Strecher et al.. This result means that Strecher's
"non-interacting solitons" had been observed by themselves. On the
other hand, under some particular initial and boundary conditions,
these solutions exactly describe the well-known collapse and
revival of a weakly interacting BEC.

We consider a BEC consisting of $N$ identical Bose atoms and being
transferred into a cigar-shaped magneto-optical trap. Dynamics of
the system is governed by the Gross-Pitaevskii equation (GPE)
\cite{{Dalfovo}}, \cite{Leggett}
\begin{eqnarray}
i \hbar \frac{\partial \psi}{\partial t}=- \frac {\hbar^2}{2m}
\bigtriangledown ^2 \psi + \Big[\frac {1}{2}m \omega_x^2 x^2 +
\frac {1}{2}m \omega_r^2(y^2+z^2) +g_0|\psi|^2\Big]\psi,
\end{eqnarray}
where $\omega_r$ and $\omega_x$ are the transverse and axial
frequencies respectively, the interaction intensity $g_0$ is
related to the $s$-wave scattering length $a$, atomic mass $m$ and
number of atoms $N$ through $g_0= 4\pi N \hbar ^2 a/m$ for the
normalized wave-function $\psi$. The norm $|\psi|^2$ is the
probability density and $N|\psi|^2$ the density of atomic number.
Setting $l_r=\sqrt{\hbar /(m \omega_r)}, \ l_x=\sqrt{\hbar /(m
\omega_x)}$ and writing $E_{kin}$ and $E_{int}$ as the kinetic
energy and mean-field interaction energy of the BEC, the
relationship $E_{int}/ E_{kin} \sim N|a|/(l_r^2l_x)^{1/3}$
expresses the importance of the atom-atom interaction compared to
the kinetic energy \cite{{Dalfovo}}. For low particle number
\cite{Wright}, \cite{Wright2} or short $s$-wave scattering length
$|a|$ \cite{Strecher}, \cite{Khaykovich}, we can treat the
interaction term as a perturbation \cite{Wright} and obtain the
leading order solution of Eq. (1) to obey the linear Schr$\ddot
o$dinger equation of a harmonic oscillator. Assuming the leading
order wave-function is in the form of separation variable
$\Psi_n=\psi_n (x, t) \psi_y (y) \psi_z (z)$ and its transverse
factor is in the ground state of harmonic oscillator, we find a
new exact solution of the linear Schr$\ddot o$dinger equation (see
Appendix A)
\begin{eqnarray}
\Psi_n &=& \psi_n (x, t) \psi_y (y) \psi_z (z)=R_n(x, y, z, t) \exp[i \Theta _n(x, t)], \nonumber \\
R_n &=& \Big[\frac{\sqrt{c_0}}{\pi \sqrt{\pi}l_r^2 l_x 2^n n! \rho
(t)}\Big]^{1/2}H_n(\xi) \exp \Big[- \frac
{1}{2}\Big(\frac{y^2}{l_r^2}+\frac{z^2}{l_r^2}+ \xi^2\Big)\Big], \
\ \xi=\frac{\sqrt{c_0}x}{\rho (t)l_x}-\frac{b_0}{\sqrt{c_0}}\cos \theta(t), \ \ \ \\
\Theta _n&=& \frac{\dot \rho (t) x^2}{2\rho (t)l_x^2}-\frac{b_0
x}{\rho (t)l_x}\sin \theta(t) +\frac{b_0^2}{4c_0}\sin [2\theta(t)]
-\Big(\frac 1 2 +n \Big)\theta(t)-\omega_r t, \ \ n=0, \ 1, \ 2,
\cdots.  \nonumber
\end{eqnarray}
Here $b_0$ is an arbitrary constant, $H_n(\xi)$ denotes the
Hermitian polynomial of variable $\xi, \ c_0= \dot \theta(t)
\rho^2(t)=AB \omega_x \sin (\alpha -\beta)$ is a conserved
quantity of the classical harmonic oscillator equation $\ddot
\varphi =-\omega_x^2\varphi$ with the complex solution $\varphi
=\rho (t) \exp[i \theta (t)], \ \rho (t)$ and $\theta (t)$ are the
amplitude and phase of the complex oscillator
\begin{eqnarray}
\rho = \sqrt{A^2 \cos^2 (\omega_x t+\alpha)+B^2 \cos ^2(\omega_x
t+\beta)}, \ \ \theta = \arctan \frac{B \cos (\omega_x t+\beta)}{A
\cos (\omega_x t+\alpha)}
\end{eqnarray}
with $A, \ B, \ \alpha$ and $\beta$ being arbitrary constants
adjusted by the initial conditions of the classical harmonic
oscillator. Obviously, the exact solution is not a energy
eigenstate, but denotes a new kind of coherent states. In
mathematical point of view, it is a complete solution with
independent constants $A, \ B, \ \alpha, \ \beta$ and $b_0$. By
adjusting these constants, we can use the exact solution to
describe some different experimental results. It is easily to
prove the exact solution (2) obeying the orthonormalization
condition (see Appendix A).

Amplitude $R_n$ of the exact solution (2) describes the
wave-packet trains consisting of $n+1$ packets. By using Eq. (3),
from $\xi =0$ we have orbit of the center of wave-packet trains
\begin{eqnarray}
x_c= \frac{b_0}{c_0}\rho (t) \cos \theta(t)=\frac{b_0}{c_0}A \cos
(\omega_x t+\alpha),
\end{eqnarray}
which is proportional to real part of the complex solution
$\varphi$ for a classical harmonic oscillator of unit mass with
amplitude $A b_0/c_0$, frequency $\omega_x$ and initial phase
$\alpha$. We call average energy of the system in state (2) as the
macroscopic quantum level, which reads (see Appendix B)
\begin{eqnarray}
E_n=\Big(\frac 1 2 +n\Big)\frac {c_1}{c_0}+\omega_r +
\frac{b_0^2}{c_0}c_2, \ \ \ \ n=0, \ 1, \ 2, \ \cdots,
\end{eqnarray}
where constant $c_1= (\dot \rho ^2 + c_0^2 / \rho^2+\rho^2
\omega_x^2)/2$ is another conserved quantity of the classical
complex oscillator (3), $b_0^2c_2/c_0=
b_0^2A^2c_1/[c_0^2(A^2+B^2)]$ is proportional to square of the
amplitude of Eq. (4) or energy of the classical oscillator. Here
and the following, we adopt the natural unit with $m=\hbar
=\omega_x =1$ such that $\omega_r$ is normalized by $\omega_x$.
For given constants, Eq. (5) exhibits that the energy only depends
on the quantum number $n$. It is quite interesting that the
average energy (5) is proportional to both the quantum level and
the classical energies of the harmonic oscillators governed by
Eqs. (3) and (4).

The function $\rho (t)$ included in $\xi$ describes not only the
total width of a wave-packet train but also the width of each
packet. The average width of the packets is $\rho(t)/\sqrt{c_0}$.
Same function in radical of Eq. (2) governs highnesses of every
packets. These and Eq. (3) infer that the widths and highnesses
may change simultaneously with frequency $2\omega$. When the
changes of the widths and highnesses are small, behavior of the
wave-packet trains seemingly to be an array of solitons. And
larger changes of the widths and highnesses show collapse and
revival of the wave-packet trains, like multiple breathers. The
normalization condition implies that the broader wave-packet train
is associated with smaller mean highness and the narrower
wave-packet train corresponds to larger mean highness. In the
process of propagation and breathing, the wave-packet trains may
spontaneously transit from state of higher average energy to that
of lower one. Some perturbations also could cause transition
between the states of different quantum numbers. All of these will
be numerically illustrated as follows.

\bf Propagation of the wave-packet trains: pure bright soliton
trains \rm A pure soliton always keeps its shape and energy in
propagation, which can be described by the exact solution (2)
through selecting $\rho$ as a constant to fix the width and
highness of each wave-packet. To do this we only require limiting
the constants in Eq. (3) to $A=B$ and $\beta =\alpha -\pi /2$ such
that $\rho =A$. Adopting such selection to the second of Eq. (3)
and to the formulas of constants $c_i$ leads to the function
$\theta (t)=\omega_x t+\alpha$ and constants $c_0=A^2 \omega_x , \
c_1=\omega_x^2A^2, \ c_2=\omega_x/2$. In this case the ground
state of Eq. (2) with $n=0$ is just the common coherent state of a
harmonic oscillator \cite{Schrodinger}-\cite{Howard}. Inserting
these constants into Eq. (5) yields the average energy
$E_n=(1/2+n)\omega_x+\omega_r+b_0^2/(2A^2)$. Final term of the
energy is just the energy of classical harmonic oscillator (4),
namely $\frac 1 2 \omega_x^2(b_0A/c_0)^2 =b_0^2/(2A^2)$.
Therefore, the average energy equates to sum of the quantum level
and classical energy. We take the parameter set $n=10, \ \alpha=0,
\ \beta =-\pi/2, \ A=B=\rho =c_0=1, \ b_0=-5, \ \omega_r=40
\omega_x$ and adopt the units of time, space and probability
density as $\omega_x^{-1}, \ l_x$ and $l_x^{-3}$ from Eq. (2) to
make the plot of $R^2$ on $xoy$ plane for different times. In 1-3
lines of Fig. 1, we show the motions of eleven pure solitons from
$t=0, \ x_c(0)=-5 l_x$ to $\omega_x t= \pi /2, \ x_c(\pi/2)=0$,
and to $\omega_x t=\pi, \ x_c(\pi)=5 l_x$. In the propagation, the
soliton train keeps its shape and distance between two solitons.

\begin{figure}[htbp]
\centering
\includegraphics[width=4.0in]{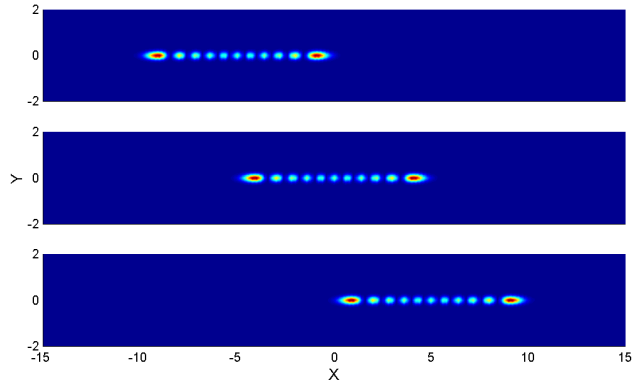}
\hspace{1cm} \vspace{-7cm} \caption{The probability density of the
wave-packet train as pure multi-solitons. Taking the parameter set
$n=10, \ \alpha=0, \ \beta =-\pi/2, \ A=B=\rho =c_0=1, \ b_0=-5, \
\omega_r=40 \omega_x$ and normalizing the time, space and
probability density in units $\omega_x^{-1}, \ l_x$ and
$l_x^{-3}$, from Eq. (2) we make the plot of $R^2$ on $xoy$ plane
for different times. In the first line of Fig. 1 the initial
profile of the soliton train is exhibited on left side of the
trap. The soliton train propagates to center of the trap at
$\omega_x t=\pi/2$ and to right side of the trap at $\omega_x
t=\pi$, as in the second and third lines of Fig. 1.}\label{fig1}
\end{figure}

\bf Breathing of the wave-packet trains: collapse and revival \rm
For a very small constant $b_0$, Eq. (4) indicates that the
amplitude of wave-packet oscillation may be very small. When the
constant $b_0$ is taken as zero, center of the wave-packet trains
is fixed to $x_c=0$ and their energy is reduce to $E_n=\Big(\frac
1 2 +n\Big)\frac {c_1}{c_0}+\omega_r$ by Eq. (5). In order to show
the collapse and revival, we let constant $B$ be much greater than
$A$, namely $B=1$ and $A=0.01$ such that the highness and width of
the wave-packet trains oscillate with greater amplitudes. The
other parameters are taken as $n=10, \ \alpha =0, \ \beta =-\pi/2,
c_0=AB=0.01, \ \omega_r/\omega_x=l_x^2/l_r^2=40$. Using the units
of space-time coordinates and probability density of Fig. 1, we
numerically draw the plots of vertical view for the wave-packet
trains as Fig. 2.

\begin{figure}[htbp]
\centering
\includegraphics[width=4.0in]{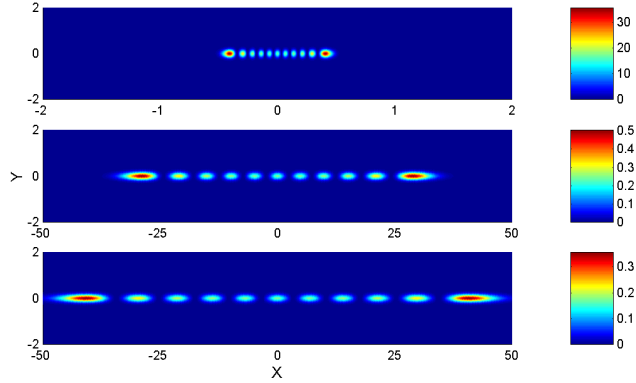}
\hspace{1cm} \vspace{-7cm} \caption{Collapse and revival of the
wave-packet trains described by Eqs. (2) and (3) with parameters
$n=10, \ b_0=0, \ A=c_0=0.01, \ B=1, \ \alpha =0, \ \beta =-\pi/2,
\ \omega_r/\omega_x=l_x^2/l_r^2=40$. The space-time coordinates
and the density are normalized in same units with Fig. 1.
Different colors represent different highnesses of the
wave-packets as in right side of each line of Fig. 2. The
initially high and narrow wave-packet train is shown in the first
line of Fig. 2. As time increases to $\omega_x t=\pi/4$, the
wave-packets collapse to highness of $10^{-1}$ order, as in the
second line of Fig. 2. To the half period $\omega_x t=\pi/2$, the
highness is reduced to $10^{-2}$ order, as in final line of Fig.
2.}\label{fig2}
\end{figure}

In Fig. 2 we exhibit that center of the wave-packet train is fixed
at $x_c=0$, width and highness of each packet are greatly changed
with period $\pi$. The first line of Fig. 2 is corresponded to the
initial wave-packet train, which is higher and narrower. The
second line of Fig. 2 shows the highnesses of packets have been
reduced to $10^{-1}$ order at $\omega_x t=\pi/4$, and total width
of the wave-packet train is simultaneously raised to $38(l_x)$.
The highness and width of the wave-packet train are collapsed to
$10^{-2}$ order and $45(l_x)$ respectively, when time equates to
the half period, as in third line of Fig. 2. In the next half
period, revival of the wave-packet train will occur, through an
inverse process of that described by Fig. 2. Fixing constant $B$,
the forms of Eqs. (2) and (3) means that highnesses of the
wave-packets are inversely proportional to constant $A$. When this
constant is taken as very small, $A\approx 0$, highness of the
wave-packet near $x=0$ will tend to infinity, resulting in the
intermittent implosions of the BEC \cite{Saito}, \cite{Hai2}.

\bf Strecher's matter-wave soliton trains \rm Generally, the
wave-packet trains governed by Eq. (2) will propagate and breathe
simultaneously, and spontaneously transit sometimes. We shall
demonstrate that these behaviors can be strictly fit to Strecher's
matter-wave soliton trains. In the experiment reported by Strecher
and coworkers \cite{Strecher}, the $^7$Li atomic BEC is employed
to create the soliton trains, by using a Feshbach resonance to
manipulate the sign and magnitude of the $s$-wave scattering
length $a$. For a small value of $|a|$ and treating the atom-atom
interaction as a perturbation proportional to $a$, the
non-interacting soliton trains may be generated.

\begin{figure}[htbp]
\centering
\includegraphics[width=2.0in]{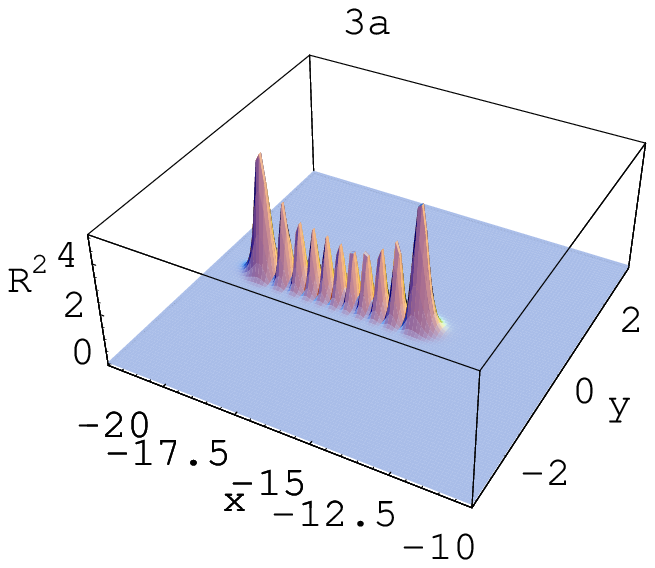}
\includegraphics[width=2.0in]{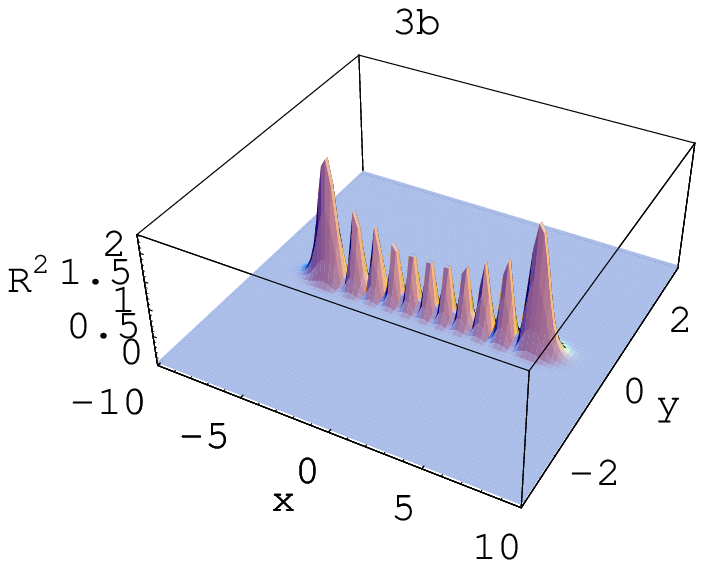}
\includegraphics[width=2.0in]{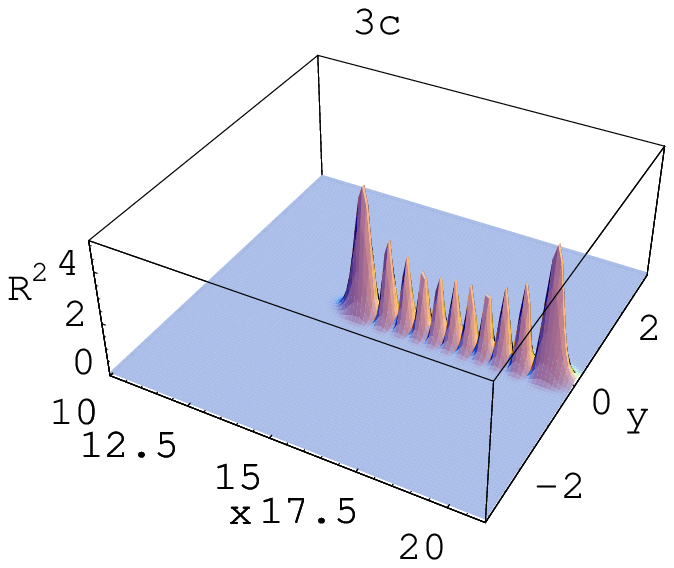}
\vspace{0.2cm} \caption{The Strecher's matter-wave soliton trains
on $xoy$ plane from Eqs. (2) and (3) with the parameters $n=10, \
A=c_0=0.4624, \ B=1, \ \alpha =0, \ \beta =-\pi/2, \ b_0=-17.437,
\ \omega_r/\omega_x=l_x^2/l_r^2=40, \ \omega_x=20$Hz,
$l_x=21.22\mu$m for (3a) $t=0$, (3b) $\omega_x t=\pi/2$ and (3c)
$\omega_x t=\pi$. The space-time coordinates and the probability
density are normalized in same units with Fig. 1. At $t=0$, center
of the soliton train is localized at
$x_c(0)=-17.437(l_x)=-370\mu$m, average width of the solitons is
$\rho(0)/\sqrt{c_0} =0.68(l_x)$, as in Fig. 3a. As time increasing
to $\omega_x t=\pi/2$, Fig. 3b shows that the center of the
soliton train arrives at $x_c(\pi/2)=0$ and the average width of
the solitons is increased to $\rho(\pi/2)/\sqrt{c_0} =1.471(l_x)$.
By Fig. 3c we display that the soliton train has moved to another
end of the trap and its shape is changed back to initial case at
$\omega_x t=\pi$. }\label{fig3}
\end{figure}

It is specially interesting to investigate the formation of the
soliton trains. Our theory shows that the non-interacting bright
soliton trains could be generated for $a \approx 0$ by the
excitation from external fields. This means that the bright
solitons described by Eq. (2) cannot be created in the case $a>0$
of Strecher's experiment. Only when the $s$-wave scattering length
changes sign from positive to negative, the generating condition
of the non-interacting solitons $a \approx 0$ can be reached.
Setting $\Delta t$ as the interval between the time the end caps
of Strecher's experiment are switched off to the time when $a$
changes sign $(a = 0)$, Strecher et al. found that number of the
solitons increases linearly with $\Delta t$. At $\Delta t=0$,
namely the case that \bf the end caps are switched off at $a=0$,
four non-interacting solitons were observed \rm \cite{Strecher}.
The larger $\Delta t$ corresponds with stronger disturbatnce and
the later can cause the system to higher excitation state with
larger mean energy given in Eq. (5). The larger quantum number $n$
is associated with more solitons. Because the known wave-packet of
a quantum harmonic oscillator does not vary its width, Strecher et
al. think of the soliton trains with variable width to be
nonlinear multi-solitons. They also expected the non-interacting
solitons being simultaneously released from different points in a
harmonic potential. Given the exact solution (2) and macroscopic
quantum level (5), the above analysis reveals that the soliton
trains found by Strecher et al. just are the non-interacting
solitons expected by themselves.

Initial number of the solitons may be greater than ten in the
experiment, oscillating frequency of the center of soliton trains
is about $20$Hz for the period $T\approx 310$ms and amplitude of
that is $370 \mu$m. This frequency was identified as the axial
one, $\omega_x$, in the previous analytical work \cite{Khawaja}.
Their radial frequency is about $40$ times the axial one. From
Fig. 4 of Ref. \cite{Strecher} we estimate that the maximum and
minimum widths of the soliton trains are about $310\mu$m and
$140\mu$m.These experimental data give limitations to the
parameters in Eqs. (2) and (3) as $\omega_x=\omega_r/40=20$Hz,
$l_x=\sqrt{40}l_r =21.22\mu$m, $Ab_0/c_0=-17.437l_x=-370 \mu$m,
$\rho(0)/\sqrt{c_0}=6.8l_x=144.30\mu$m,
$\rho(\pi/2)/\sqrt{c_0}=14.71 l_x=312.15\mu$m. Under these
limitations, we choose the parameter set $n=10, \ A=c_0=0.4624, \
B=1, \ \alpha =0, \ \beta =-\pi/2, \ b_0=-17.437$ and use the
units of space-time coordinates and probability density of Fig. 1
to make 3D plots of the soliton train for the time $\omega_x t=0,
\ \pi/2$ and $\pi$, respectively, as Fig. 3a, 3b and 3c. These
plots display that the soliton train localized at ends of the trap
possesses minimum width. When it move to center of the trap, its
width becomes maximum. The change of the width was explained as
repulsive interaction among the solitons in previous work
\cite{Strecher}, \cite{Khawaja}. In motion of the soliton train,
its highness has only small change such that effect of the
collapse and revival cannot be observed. The solitons at ends of
any soliton train are higher and thicker compared to the other
ones. The thickness of each soliton is shown in Fig. 4, which is
the vertical view of Fig. 3. This graph is very like the figure 4
of Strecher' article \cite{Strecher}, so the former could be a
good fit to the latter. The differences of highness and thickness
have not been accurately distinguished in the previous experiment.
If this is done, we can fit it better by applying Eq. (2) and
adjusting the constants $A, \ B, \ \alpha, \ \beta$ and $b_0$.

\begin{figure}[htbp]
\centering
\includegraphics[width=4.0in]{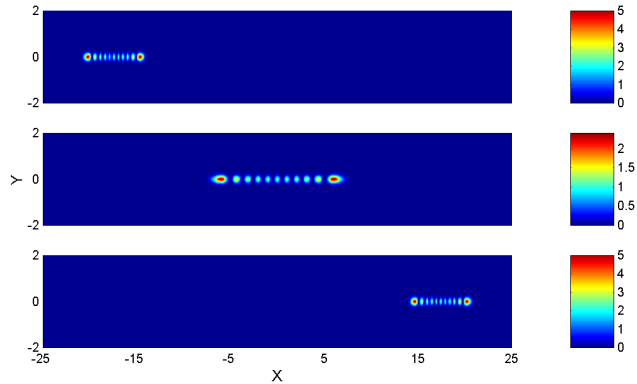}
\hspace{1cm} \vspace{-7cm} \caption{The vertical view of Fig. 3
with different line being corresponded to Fig. 3a, 3b and 3c
respectively. Comparison between this with the figure 4 of
Strecher's article exhibits good agreement between
them.}\label{fig4}
\end{figure}

\begin{figure}[htbp]
\centering
\includegraphics[width=4.0in]{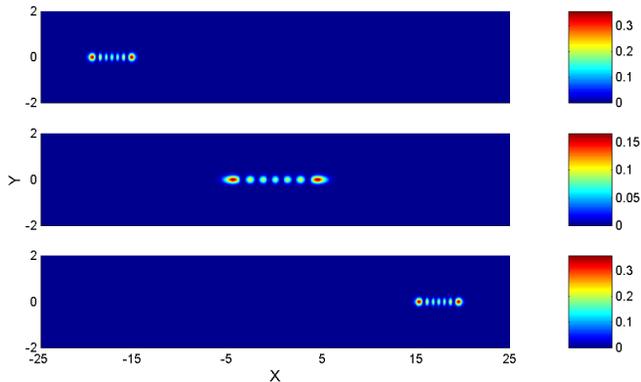}
\hspace{1cm} \vspace{-7cm} \caption{Transition of the matter-wave
soliton trains. If the state of energy $E_{10}$ is disturbed by
the interatomic interaction, the spontaneous transitions to the
states of less average energy could occur. In Fig. 5 we show that
the train consisting of eleven solitons is transformed into that
of seven solitons for $n=6$. The different lines correspond to the
times $\omega_x t=2\pi, \ 2.5\pi$ and $3\pi$
respectively.}\label{fig5}
\end{figure}

\bf Spontaneous transitions of the wave-packet trains \rm In
Strecher's experiment on the matter-wave solitons, the trains with
missing solitons were frequently observed, and this is resided in
loss of condensed atoms. However, "it is not clear whether this is
because of a slow loss of atoms, or because of sudden loss of an
individual soliton" \cite{Strecher}. According to transition
theory of quantum-mechanical states, the non-interacting solitons
could transit from higher-energy state to lower-energy state. The
state function (2) and energy (5) imply that the lower-energy
state describes less solitons. Therefore, even if the condensed
atoms propagate without loss of number, some solitons may be
suddenly lost, through spontaneous transitions of the macroscopic
quantum states. Just after a transition, of course, the highness
and width of each soliton should increase, since the atoms in lost
solitons have entered the remainder solitons. The spontaneous
transitions may be random and can be caused by some perturbations.
We assume that under perturbation of the interatomic interaction
the state of eleven solitons in Fig. 3 transits to the states with
$n=6$ at $\omega_x t=2 \pi$ as in the first line of Fig. 5 and
propagates to $\omega_x t=2.5 \pi$ and $\omega_x t=3 \pi$ as in
the second and third lines of Fig. 5.

Further considering the first order correction to the wave-packet
trains from weak interaction and investigating transformation of
the interaction from weak to strong will be very interesting.
Because of the existence of arbitrary constants $A, \ B, \ \alpha,
\ \beta, \ b_0$ and periodic functions $\rho(t), \ \theta(t)$, by
using the complete solution (2) we can control motions of the BEC
wave-packets. The theoretical control could indicate the
directions of experimental operations that is important for real
application, say, making an atomic soliton laser based on the
bright soliton trains. In addition, the exact solution (2) could
play an important role in treating various harmonically confined
systems. For example, a single Paul trapped ion interacting with a
harmonic potential, the state $\Psi_1$ of two wave-packets is
similar to the Schr$\ddot o$dinger's cat state \cite{Monroe},
\cite{Brune}.

\bf Acknowledgments \rm  This work was supported by the NNSF of
China under Grant No. 10275023 and the NLMRAMP of China under
Grant No. T152103, and by the Hubei Provincial Key Laboratory of
Gravitation and Quantum Physics of China.

\newpage

\begin{center}
\bf Appendix A: \ Derivation of the Exact Solution
\end{center}

\rm We adopt the natural unit with $m=\hbar =\omega_x =1$ and
insert $\Psi_n=\psi_n (x, t) \psi_y (y) \psi_z (z)=
(\sqrt{\pi}l_r)^{-1} \exp [-(y^2+z^2)/(2l_r^2)]\psi_n (x, t), \
a=0$ into Eq. (1), producing the one dimensional equation of a
harmonic oscillator
\begin{eqnarray}
i \hbar  \frac{\partial \psi_n}{\partial t}=- \frac {1}{2}
\frac{\partial^2 \psi_n}{\partial x^2} + \Big[\frac {1}{2}
\omega_x^2 x^2 + \omega_r \Big]\psi_n, \ \ \ \ \ \ \ \ \ \ \ \ \ \
\ \ \ \ \ \ \ \ \ \ \ (A1) \nonumber
\end{eqnarray}
where units of $t$ and $x$ are $\omega_x^{-1}$ and $l_x$
respectively, $\omega_r$ is normalized by $\omega_x$ and
appearance of the later is only formal, since its value has been
fixed to 1. Let the solution of Eq. (A1) be in the form
\begin{eqnarray}
\psi_n=a_n(t)H_n(\xi)\exp [b(t)x-c(t)x^2-f^2(t)/2], \ \
\xi=e(t)x-f(t) \ \ \ \ \ (A2) \nonumber
\end{eqnarray}
with $a(t), \ b(t), \ c(t)$ being the complex functions of time
and $e(t), \ f(t)$ the real functions. Applying Eq. (A2) to Eq.
(A1), we arrive at the equation
\begin{eqnarray}
e^2 \frac{\partial^2 H_n}{\partial \xi^2}&+&2(be-i\dot f+i\dot e
x-2cex)\frac{\partial H_n}{\partial \xi} \nonumber
\\ + 2\Big[i\frac{\dot {a}_n}{a_n}&-& i f\dot f+\frac{b^2}{2}-c-
\omega_r+(i\dot b-2bc)x+\Big(2c^2-i\dot c-\frac 1 2 \omega_x^2
\Big)x^2\Big]H_n=0. \ \ \ (A3) \nonumber
\end{eqnarray}
Noticing the Hermitian equation $\partial^2 H_n / \partial \xi^2
-2\xi \partial H_n / \partial \xi +2nH_n =0$, Eq. (A3) implies
\begin{eqnarray}
i \dot c &=&2c^2-\omega_x^2/2, \ \ i \dot b =2bc, \ \ i \dot e
=2ce-e^3, \nonumber \\  i \dot f &=&be- e^2f, \ \ i \dot {a}_n
/a_n =i f\dot f -b^2/2+c+\omega_r+ne^2. \ \ \ \ \ \ \ \ \ \ (A4)
\nonumber
\end{eqnarray}
The first of Eq. (A4) is a complex Riccati equation, which can be
changed into a complex equation of a classical harmonic oscillator
\begin{eqnarray}
\ddot \varphi =-\omega_x^2 \varphi, \ \ \ \ \ \ \ \ \ \ \ \ \ \ \
\ \ \ \ \ \ \ (A5) \nonumber
\end{eqnarray}
through the function transformation $c=\dot \varphi /(2i
\varphi)$. The general solution of Eq. (A5) is well-known that
\begin{eqnarray}
\varphi =A \cos (\omega_x t+\alpha)+i B\cos (\omega_x
t+\beta)=\rho(x, t)\exp[i \theta (x, t)], \ \ \ \ \ \ \ \ \ \ \ \
\ \ \ \ \ \ (A6) \nonumber
\end{eqnarray}
where $A, \ B, \ \alpha$ and $\beta$ are arbitrary constants, the
real functions $\rho (x, t)$ and $\theta (x, t)$ have been given
by Eq. (3). Returning to the transformation between $\varphi$ and
$c$ yields
\begin{eqnarray}
c=\frac{\dot \varphi}{2i \varphi}= \frac 1 2 \dot \theta -i
\frac{\dot \rho}{2\rho}. \ \ \ \ \ \ \ \ \ \ \ \ \ \ \ \ \ \ \ \ \
\ (A7) \nonumber
\end{eqnarray}
Substitution of Eq. (A6) into Eq. (A5) yields equations of the
amplitude and phase as
\begin{eqnarray}
\ddot \theta =-2\dot \theta \dot \rho / \rho, \ \ \  \ddot
\rho=\rho \dot \theta ^2 -\omega_x^2\rho \ \ \ \ \ \ \ \ \ \ \ \ \
\ \ \ \ \ \ \ \ \ (A8) \nonumber
\end{eqnarray}
with the first integrations
\begin{eqnarray}
c_0=\rho^2 \dot \theta =AB \omega_x \sin (\alpha -\beta), \ \ \ \
c_1= (\dot \rho ^2 + c_0^2 / \rho^2+\rho^2 \omega_x^2)/2. \ \ \ \
\ \ \ \ \ \ \ \ \ \ \ \ \ \ \ \ \ \ (A9) \nonumber
\end{eqnarray}
Combining Eqs. (A7) and (A6) with Eq. (A4) and applying the
relation (A9), we easily obtain
\begin{eqnarray}
b&=&b_0 \frac{\exp(-i\theta)}{\rho}, \ e=\frac{\sqrt{c_0}}{\rho}
=\sqrt{\dot \theta}, \ f=\frac{b_0}{\sqrt{c_0}}\cos \theta ,
\nonumber \\ a_n&=&\frac{A_0}{\sqrt{\rho}}\exp
\Big(-i\Big[\Big(\frac 1 2 +n\Big)\theta +\omega_r t
-\frac{b_0^2}{4c_0}\sin 2\theta\Big]\Big). \ \ \ \ \ \ \ \ \ (A10)
\nonumber
\end{eqnarray}
Inserting these into Eq. (A2) leads to the axial solution
$\psi_n(x, t)$, and the normalization condition $\int |\psi_n|^2
dx=A_0^2\sqrt{c_0^{-1}}\int
H_n^2(\xi)\exp(-\xi^2)d\xi=A_0^2\sqrt{\pi c_0^{-1}}2^n n!=1$ gives
the constant $A_0=[\sqrt{c_0}/(\sqrt{\pi}2^n n!)]^{1/2}$ such that
Eq. (A2) becomes
\begin{eqnarray}
\psi_n(x, t)= \Big[\frac{\sqrt{c_0}}{\sqrt{\pi} 2^n n! \rho
(t)}\Big]^{1/2}H_n(\xi) \exp \Big[-\frac {1}{2} \xi^2 +i\Theta_n
(x, t)\Big],\ \ \ \ \ \ \ \ \ (A11) \nonumber
\end{eqnarray}
where the function $\Theta(x, t)$ has been written in Eq. (2).
Combining this with the wave-function of transverse dimension and
letting the unit of spatial coordinate return to original one,
$x\rightarrow x/l_x$, we finally get the normalized 3D
wave-function, as in Eq. (2).

\begin{center}
\bf Appendix B: \ Proof of the Average Energy
\end{center}

Employing the Dirac's symbols, ket and bra, from Eq. (A2) and the
quantum-mechanical definition of average energy in state $\psi_n$
we perform the calculation
\begin{eqnarray}
E_n=\langle \psi_n|i\frac {\partial}{\partial t}\psi_n\rangle =i
\langle \psi_n|\frac{\dot a_n}{a_n}+(\dot e x-\dot f)\frac{1}{H_n}
\frac{\partial H_n}{\partial \xi}-(f\dot f-\dot b x+\dot c
x^2)|\psi_n\rangle . \ \ \ \ \ \ \ \  (A12) \nonumber
\end{eqnarray}
Noticing the orthonormalization condition $\langle
\psi_n|\psi_{n'}\rangle =\delta_{nn'}$ and the formulas
\begin{eqnarray}
\xi=e(t)x-f(t), \ \ \xi \psi_n=
\sqrt{n/2}\psi_{n-1}+\sqrt{(n+1)/2}\psi_{n+1}, \ \nonumber \\
2\xi^2
\psi_n=\sqrt{n(n-1)}\psi_{n-2}+(2n+1)\psi_n+\sqrt{(n+1)(n+2)}\psi_{n+2}
, \nonumber
\\
\frac {1}{H_n} \frac{\partial H_n}{
\partial \xi
}\psi_n=2n\frac{H_{n-1}}{H_n}\psi_n=\sqrt{2n}\psi_{n-1}, \ \ \ \ \
\ \ \ \ \ \ \ \ \ \nonumber
\end{eqnarray}
we continue to compute the average energy
\begin{eqnarray}
E_n&=& i\Big(\frac{\dot a_n}{a_n}-f\dot f\Big)+i\langle
\psi_n|\sqrt{2n}\dot e \xi /e|\psi_{n-1}\rangle+i\langle
\psi_n|\dot b x-\dot c x^2|\psi_{n}\rangle \nonumber \\
&=& i\Big(\frac{\dot a_n}{a_n}-f\dot f+n\frac{\dot
e}{e}-\frac{\dot c}{e^2}\Big[f^2+\Big(\frac 1 2
+n\Big)\Big]\Big)+i\frac{\dot b f}{e}.   \nonumber
\end{eqnarray}
Applying Eqs. (A4) to the above equation results in
\begin{eqnarray}
E_n &=&
(2n+1)c+\omega_r-\frac{b^2}{2}-\Big(2c^2-\frac{\omega_x^2}{2}\Big)\Big[f^2+\Big(\frac
1 2 +n\Big)\Big]\frac{1}{e^2}+2bc\frac{f}{e}.  \ \ \ \ \ \ \ \ \ \
\ \ \ \ (A13) \nonumber
\end{eqnarray}
Noticing Eqs. (A7), (A9) and (A10) we have
\begin{eqnarray}
(2n+1)c-\Big(2c^2-\frac{\omega_x^2}{2}\Big)\Big(\frac 1 2
+n\Big)\frac{1}{e^2}=\Big(\frac 1 2 +n\Big)\frac {c_1}{c_0}, \ \
2bc\frac{f}{e}-\frac{b^2}{2}-\Big(2c^2-\frac{\omega_x^2}{2}\Big)\frac{f^2}{e^2}=\frac{b_0^2}{c_0}c_2,
\ \nonumber \\  c_2=\frac{\dot
\theta}{2}+\Big(\frac{c_1}{c_0}-\dot \theta \Big)\cos^2 \theta
-\frac{\dot \rho}{\rho}\cos \theta \sin \theta. \ \ \ \ \ \ \ \ \
\ \ \ \ \ \ \ \ \ \ \ \ \ \ \ \ \ \nonumber
\end{eqnarray}
The third equation and Eqs. (A6) and (A9) imply
\begin{eqnarray}
\frac{c_2}{\dot \theta}&=&\frac{c_2}{c_0}\rho^2=\frac 1 2 -\cos^2
\theta+\frac{c_1}{c_0^2}(\rho \cos \theta)^2-\frac{1}{c_0}(\rho
\sin \theta) \dot \rho \cos \theta \nonumber \\ &=&\frac 1 2
-\cos^2 \theta+\frac{c_1}{c_0^2}(\rho \cos
\theta)^2-\frac{1}{c_0}(\rho \sin \theta)\Big[\frac{c_0}{\rho}\sin
\theta-A\omega_x\sin(\omega_x t+\alpha)\Big] \nonumber
\end{eqnarray}
so that we get
\begin{eqnarray}
\frac{c_2}{c_0}[A^2 \cos^2 (\omega_x t+\alpha)+B^2 \cos
^2(\omega_x t+\beta)] \nonumber  \ \ \ \ \ \ \ \ \ \ \ \ \ \ \ \ \ \ \ \ \ \ \ \ \ \   \\
=-\frac 1 2 +\frac{c_1}{c_0^2}A^2\cos^2(\omega_x
t+\alpha)+\frac{\omega_x}{c_0}AB\cos (\omega_x t+\beta)\sin
(\omega_x t+\alpha). \nonumber
\end{eqnarray}
Let the function of time be in the forms $\sin (2\omega_x t)$ and
$\cos (2\omega_x t)$, and identify the corresponding coefficients
of both sides, producing
\begin{eqnarray}
c_2 = \frac{A^2c_1}{(A^2+B^2)c_0}. \nonumber
\end{eqnarray}
Combining these with Eq. (13) leads to
\begin{eqnarray}
E_n=\Big(\frac 1 2 +n\Big)\frac {c_1}{c_0}+\omega_r +
\frac{b_0^2}{c_0}c_2, \ \ \ \ n=0, \ 1, \ 2, \ \cdots. \nonumber
\end{eqnarray}
This is just the macroscopic quantum level (5).

\end{document}